\documentclass[aps,prl,twocolumn,showpacs,preprintnumbers,superscriptaddress,nofootinbib]{revtex4-1}

\usepackage{amsmath}
\usepackage{amssymb}
\usepackage{graphicx}
\usepackage{dcolumn}
\usepackage{bm}
\usepackage[usenames,dvipsnames]{color}
\newcommand{\BMS}{\mbox{\rm BaMn$_{2}$As$_{2}$ }}
\newcommand{\BMSend}{\mbox{\rm BaMn$_{2}$As$_{2}$}}

\newcommand{\KKBMS}{\mbox{\rm K$_{0.02}$Ba$_{0.98}$Mn$_{2}$As$_{2}$ }}

\begin{document}

\title{From Hund's insulator to Fermi liquid: optical spectroscopy study of K-doping in BaMn$_2$As$_2$}%

\author{D.E. McNally}
\email[]{daniel.mcnally@stonybrook.edu}
\affiliation{Department of Physics and Astronomy, Stony Brook University, Stony Brook, New York 11794-3800, USA}
\author{S. Zellman}
\affiliation{Department of Physics and Astronomy, Stony Brook University, Stony Brook, New York 11794-3800, USA}
\author{Z. P. Yin}
\affiliation{Department of Physics and Astronomy, Rutgers University, Piscataway, NJ 08854, USA}
\author{K.W. Post}
\affiliation{Department of Physics, University of California, San Diego, La Jolla, CA 92093-0319, USA}
\author{Hua He}
\affiliation{Department of Physics and Astronomy, Stony Brook University, Stony Brook, New York 11794-3800, USA}
\author{K. Hao}
\affiliation{Department of Physics, University of California, San Diego, La Jolla, CA 92093-0319, USA}
\author{G. Kotliar}
\affiliation{Department of Physics and Astronomy, Rutgers University, Piscataway, NJ 08854, USA}
\author{D. Basov}
\affiliation{Department of Physics, University of California, San Diego, La Jolla, CA 92093-0319, USA}
\author{C. C. Homes}
\affiliation{Condensed Matter Physics and Materials Science Department, Brookhaven National Laboratory, Upton, New York, 11973-5000, USA}
\author{M.C. Aronson}
\affiliation{Department of Physics and Astronomy, Stony Brook University, Stony Brook, New York 11794-3800, USA}
\affiliation{Condensed Matter Physics and Materials Science Department, Brookhaven National Laboratory, Upton, New York, 11973-5000, USA}

\date{\today}%

\begin{abstract}
We present optical transmission measurements that reveal a charge gap of 0.86 eV in the local moment antiferromagnetic insulator \BMS, an order of magnitude larger than previously reported. Density functional theory plus dynamical mean field theory (DFT+DMFT) calculations correctly reproduce this charge gap only when a strong Hund's coupling is considered. Thus, \BMS is a member of a wider class of Mn pnictide compounds that are Mott-Hund's insulators. We also present optical reflectance for metallic 2\% K doped \BMS that we use to extract the optical conductivity at different temperatures. The optical conductivity $\sigma_1$($\omega$) exhibits a metallic response that is well described by a simple Drude term. Both $\sigma$($\omega$$\rightarrow$0, T) and $\rho$(T) exhibit Fermi liquid temperature dependencies.  From these measurements, we argue that a more strongly correlated Hund's metal version of the parent compounds of the iron pnictide superconductors has not yet been realized by doping this class of Hund's insulators.
\end{abstract}

\pacs{Enter}

\maketitle

The discovery of high temperature superconductivity (HTSC) upon doping the parent compounds of metallic Fe-based materials challenged the validity of a Mott insulator picture as an appropriate starting point for an explanation of HTSC phenomena~\cite{imada1998,johnson2015}. It was soon recognized that a multi-orbital picture must be adopted and the correlations in Fe-based compounds arise from Hund's coupling that promotes a correlated metallic state away from half filling~\cite{haule2009, georges2013}. These observations led to extensive investigations of the half-filled layered Mn pnictide systems where Hund's coupling promotes an insulating state with a charge gap on order of 1 eV~\cite{mcnally2014, mcnally2015}.  By analogy with the cuprates, it was hypothesized that doping these Hund's insulators across a metal-insulator transition, with a concommitant suppression of the magnetic order, could lead to a superconducting phase.   

Layered insulating Mn pnictide systems like \mbox{LaMnPO}~\cite{mcnally2014} and \BMS~\cite{singh2009} that adopt the ZrCuSiAs and ThCr$_2$Si$_2$ structure types have been the subjects of the most thorough investigations, as they are isostructural with the parent compounds of the Fe-based HTSC. Significant charge fluctuations have been reported in the antiferromagnetic insulator \mbox{LaMnPO}, as evidenced by the ordered moment of 3.2 $\mu_B$, substantially reduced from the high spin value of 5 $\mu_B$ expected from Hund's rules~\cite{simonson2012}. However, the direct charge gap of 1 eV and the ordered moment are only marginally reduced by the introduction of as much as 28\% fluorine in LaMnPO$_{1-x}$F$_x$~\cite{simonson2011}. Here, the stability of the gap to chemically doping has been ascribed to the strength of Hund's coupling, in agreement with density functional theory plus dynamical mean field theory (DFT+DMFT) calculations~\cite{mcnally2014}. Chemical doping of the antiferromagnetic insulator \BMS~\cite{singh2009} met with even less success, with initial reports that limited amounts of most dopants could be incorporated into the structure with only marginal effects on the electronic properties~\cite{pandey2011}. This was perhaps unsurprising as \BMS has a much larger ordered moment (3.9 $\mu_B$/Mn) and ordering temperature (T$_N$ = 625 K)~\cite{singh2009_2} than LaMnPO, suggestive of weaker charge fluctuations and a more robust charge gap. However, the direct charge gaps in \BMS reported from optical reflectance (0.024 eV~\cite{antal2012}), ARPES (0.15 eV~\cite{pandey2012}) and DFT (0.1 eV~\cite{an2009}, 0.058 eV~\cite{pandey2012}), are an order of magnitude smaller than those found in related Mn pnictide systems such as LaMnXO (X = P, As, Sb)~\cite{kayanuma2009}. These results suggest \BMS could be unique amongst Mn pnictide systems, prompting our optical transmission and reflectance measurements that provide a direct measurement of the charge gap. Further, it was reported from electrical resistivity measurements that the introduction of only 2\% K in \BMS is sufficient to drive a metal-insulator transition and reveal an antiferromagnetic local moment metallic state with T$_N$ almost unaffected~\cite{pandey2012}. 

We present optical transmission and reflectance measurements and DFT+DMFT calculations to clarify the origin of the charge gap in undoped and insulating \BMSend, as well as the nature of the metallic state induced by 2\% K doping. Our optical transmission measurements determine a charge gap of 0.86 eV in undoped \BMSend, an order of magnitude larger than previous reports~\cite{antal2012} and very similar to values found in other square-net Mn compounds~\cite{kayanuma2009}. DFT+DMFT calculations correctly reproduce the measured charge gap only when a large J$_H$ = 0.9 eV is included, as well as Hubbard U = 8 eV. We also present optical reflectance measurements on a 2\% K doped single crystal that find metallic behavior characterized by the appearance of a Drude peak. The optical conductivity $\sigma(\omega)$ is extracted from the reflectance by Kramers-Kronig analysis and we find that $\sigma_1(\omega)$ is well described by a single Drude term. The quadratic temperature T-dependence of both the electrical resistivity and $\sigma_1(\omega\rightarrow 0)$ is as expected for a Fermi liquid for T = 1.8 K - 300 K. 



We grew single crystals of \BMS and 2\% K doped \BMS from a Sn flux as detailed elsewhere~\cite{singh2009}. The presence of 2\% K was confirmed by energy dispersive x-ray spectroscopy using a SEM JEOL 7600F, and the ThCr$_2$Si$_2$ structure~\cite{brechtel1978} was confirmed by x-ray diffraction using a Bruker D8 Advance. Electrical resistivity measurements were performed using a Quantum Design Physical Properties Measurement System. Optical reflectance measurements were performed on Bruker IFS 113v and Vertex 80v Fourier transform spectrometers.  Reflectance was measured on polished single crystal samples at a near-normal angle of incidence from ~2 meV to 3 eV using an in situ overcoating (overfilling) technique~\cite{homes93}. Infrared transmission measurements were carried out on single crystals of \BMS using a Bruker LUMOS FT-IR Microscope with a KBr window. These data were normalized to an open channel by keeping all other conditions the same but moving the sample out of the way.  

The electronic structure of \BMS was determined using DFT + DMFT~\cite{DMFT-RMP2006,Haule-DMFT}, which is based on the full-potential linear augmented plane wave method implemented in Wien2K~\cite{wien2k}, using the generalized gradient approximation to the exchange-correlation functional~\cite{perdew1996}. We use the atomic positions taken from the experimentally determined crystal structure~\cite{brechtel1978}. The convergence of the calculations with respect to number of k points, charge density, total energy, Fermi level, self-energy reached a level similar to previous publications ~\cite{yin2011_2,simonson2012}.

\begin{figure}[[htbp!]
\includegraphics[width=8.6 cm]
{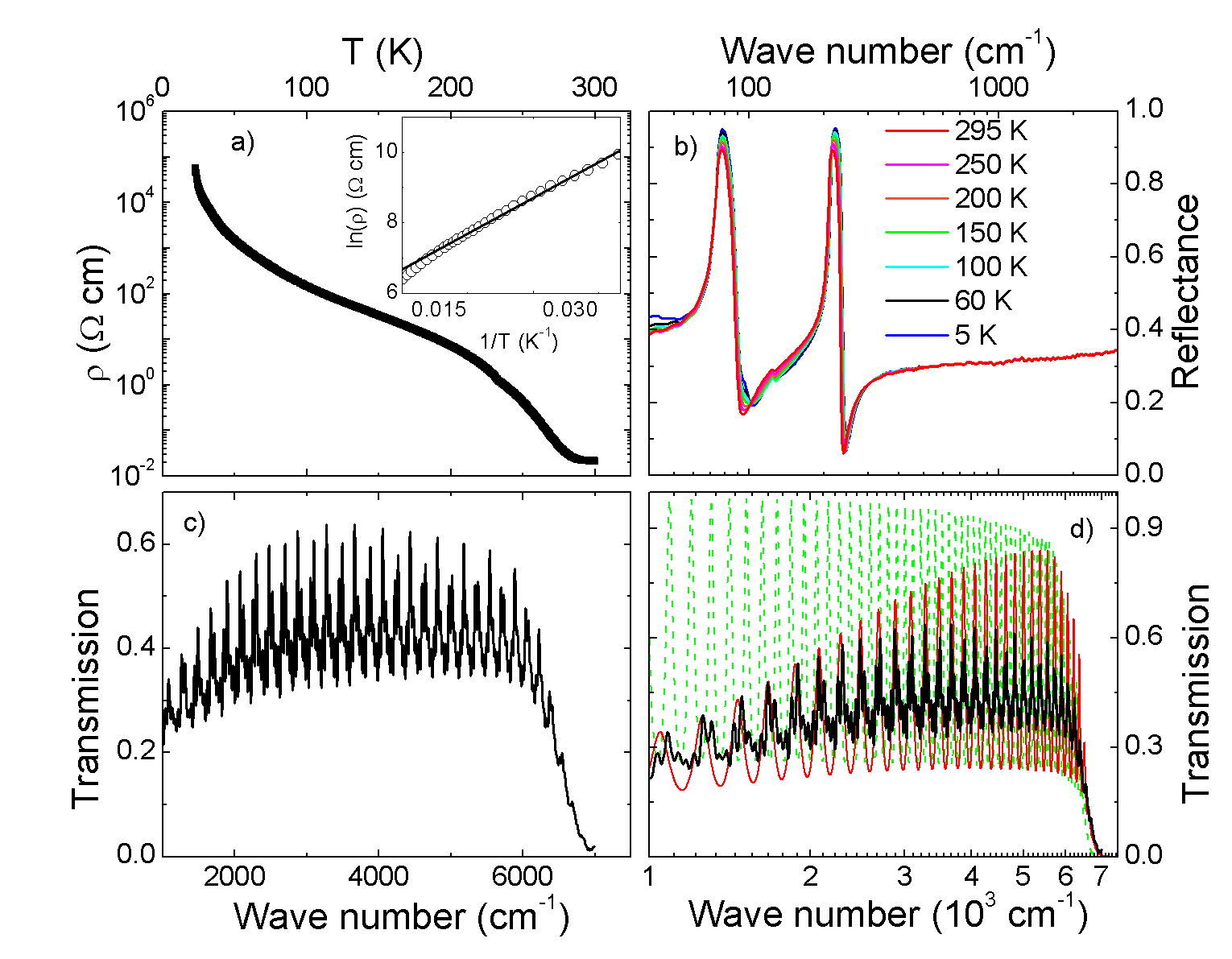}
\caption[]{(Color online) (a) Resistivity $\rho$ of BaMn$_2$As$_2$ as function of temperature T measured in the ab plane. (inset) ln$(\rho$) versus T$^{-1}$. The solid line is a fit to the Boltzmann expression as described in the main text (b) Optical reflectance measured in the ab plane for temperatures T = 295 K (red), 250 K (magneta), 200 K (orange), 150 K (green), 100 K (cyan), 60 K (black) and 5 K (blue). (c) Optical transmission measured in the ab plane at room temperature. (d) Optical transmission measured in the ab plane at room temperature (black) and a model of the optical transmission with (red) and without (green dash) intra-gap absorption.}
\label{Fig1}
\end{figure}

Figure \ref{Fig1} shows that \BMS is an insulator. The electrical resistivity $\rho$(T) measured with the current in the ab plane is presented in Figure \ref{Fig1}a and is found to increase with decreasing temperature T. The inset shows that at low temperatures $\rho$(T) is well descibed by activated T-dependence $\rho$  $\propto$ exp($\epsilon$$_A$/ k$_B$T), with an activation gap $\epsilon_A$ = 13 meV, in reasonable agreement with previous reports~\cite{singh2009}. Figure 1b presents high resolution optical reflectance of a single crystal of BaMn$_2$As$_2$ as a function of wavenumber for T = 6 K - 295 K. The reflectance is as expected for an insulator and exhibits little T-dependence. We have not been able to reproduce the T-dependent reflectance or the broad hump centered at ≈ 500 cm$^{-1}$ previously reported [12]. We emphasize that we used single crystals for our optical spectroscopy and $\rho$(T) (Fig 1a) measurements that were taken from the same batch. Room temperature optical transmission as a function of wavenumber is presented in Figure \ref{Fig1}c. At $\approx$ 6500 cm$^{-1}$ a rapid decrease of transmission is observed, consistent with the onset of absorption due to optical excitations across a charge gap $\Delta$ = 0.86 eV, significantly larger than the previously reported charge gap of 0.024 eV~\cite{antal2012}. Figure \ref{Fig1}d shows that the reduced transmission below $\approx$ 3000 cm$^{-1}$ is well accounted for by a model of Kramers-Kronig oscillators including some small intra-gap absorption due to in gap states or Sn inclusions~\cite{kuzmenko2005}. 

\begin{figure}[[htbp!]
\includegraphics[width=8.6 cm]
{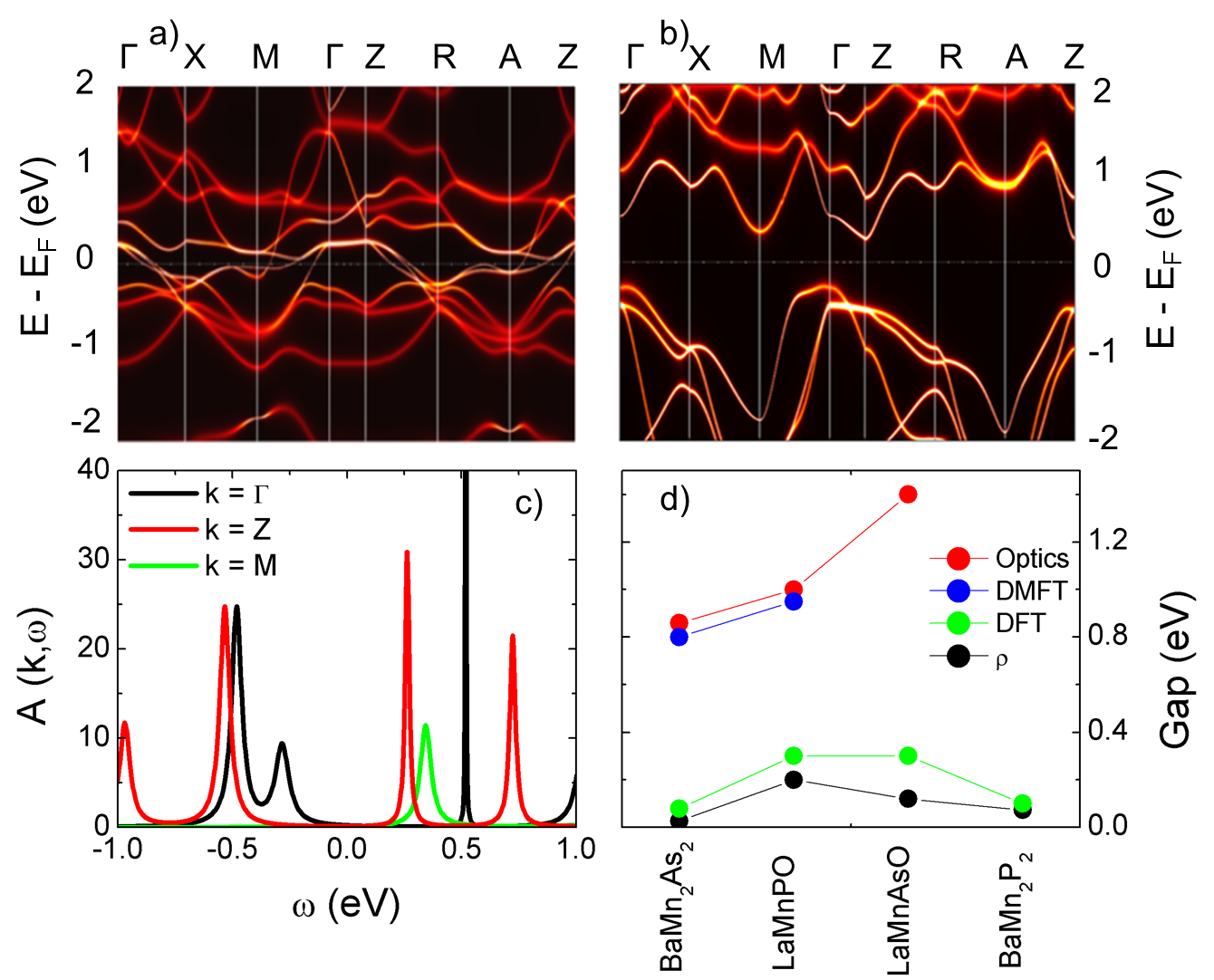}
\caption[]{(Color online) (a) Density functional theory + dynamical mean-field theory (DFT+DMFT) calculation of the band structure of \BMS with Hubbard U = 8 eV and no Hund's coupling (J$_H$ = 0 eV). (b) DFT+DMFT of the band structure of \BMS with U = 8 eV and J$_H$ = 0.9 eV. (c) Spectral function at high symmetry points for the calculation shown in (b). (d) Summary of the activation gaps $\epsilon_A$ determined from electrical resistivity $\rho$ measurements and charge gap $\Delta$ determined from different experimental and theoretical approaches in LaMnXO, BaMn$_2$X$_2$ (X = P, As)~\cite{mcnally2014, an2009, kayanuma2009,brock1994}.}
\label{Fig2}
\end{figure}

Figure \ref{Fig2} shows that the charge gap in \BMS results from strong Hund's coupling. First, in Figure \ref{Fig2}a, we present a DFT+DMFT calculation for Hubbard U = 8 eV with no Hund's term (J$_H$ = 0). Several bands are crossing the Fermi level, and so BaMn$_2$As$_2$ is predicted to be metallic, contrary to our resistivity and optical spectroscopy measurements that find insulating behavior. Figure \ref{Fig2}b presents DFT+DMFT calculations for Hubbard U = 8 eV and J$_H$ = 0.9 eV, showing that the inclusion of Hund's coupling severely modifies the band structure. Since no bands cross the Fermi level, \BMS is found to be an insulator in agreement with experiments. Further, a direct charge gap of $\approx$ 0.8 eV (Figure \ref{Fig2}c) is found at the $\Gamma$ point that is in excellent agreement with the experimental value $\Delta$ = 0.86 eV found from our optical measurements presented in Figure \ref{Fig1}.

Using our experimental and theoretical results, we now situate \BMS amongst archetypal layered Mn pnictide compounds LaMnXO and BaMn$_2$X$_2$ (X = P, As) (Figure \ref{Fig2}d). First, we discuss the experimentally determined activation gaps $\epsilon_A$ and charge gaps $\Delta$. The $\epsilon_A$ determined from electrical resistivity measurements in the ab plane range from 0.027 - 0.2 eV, while the optical gaps range from $\Delta$ = 0.86 - 1.4 eV. Thus, $\epsilon_A$ likely corresponds to the energy differences between in-gap states, possibly from impurities, and the conduction band edge. In contrast, $\Delta$ corresponds to the direct charge gap that separates the valence and conduction bands. Spin polarized DFT calculations find a direct charge gap of similar magnitude to the activation gaps, which is order of magnitude smaller than the experimentally determined charge gaps. On the other hand, DFT+DMFT calculations including a strong Hund's coupling correctly reproduce the measured charge gap. These calculations assert that Hund's coupling is crucial to properly understand the insulating state of Mn pnictide systems~\cite{mcnally2014, gibson2015}. It has also been emphasized that Hund's coupling is responsible for the mass enhancement observed in the parent compounds of the iron pnictide superconductors~\cite{yin2011_2}. Thus, driving the strongly correlated Mn pnictide Hund's insulators across an electronic delocalization transition where $\Delta$$\rightarrow$0 could potentially lead to a more correlated Mn version of the Fe pnictides~\cite{yi2013}, and perhaps even higher temperature superconductivity. While a metal-insulator transition has been realized in K-doped BaMn$_2$As$_2$, this sample was not found to be superconducting~\cite{pandey2012}. We now present $\rho(T)$ and optical spectroscopy measurements that address the metallization of 2\% K doped \BMS.

\begin{figure}[[htbp!]
\includegraphics[width=8.6 cm]
{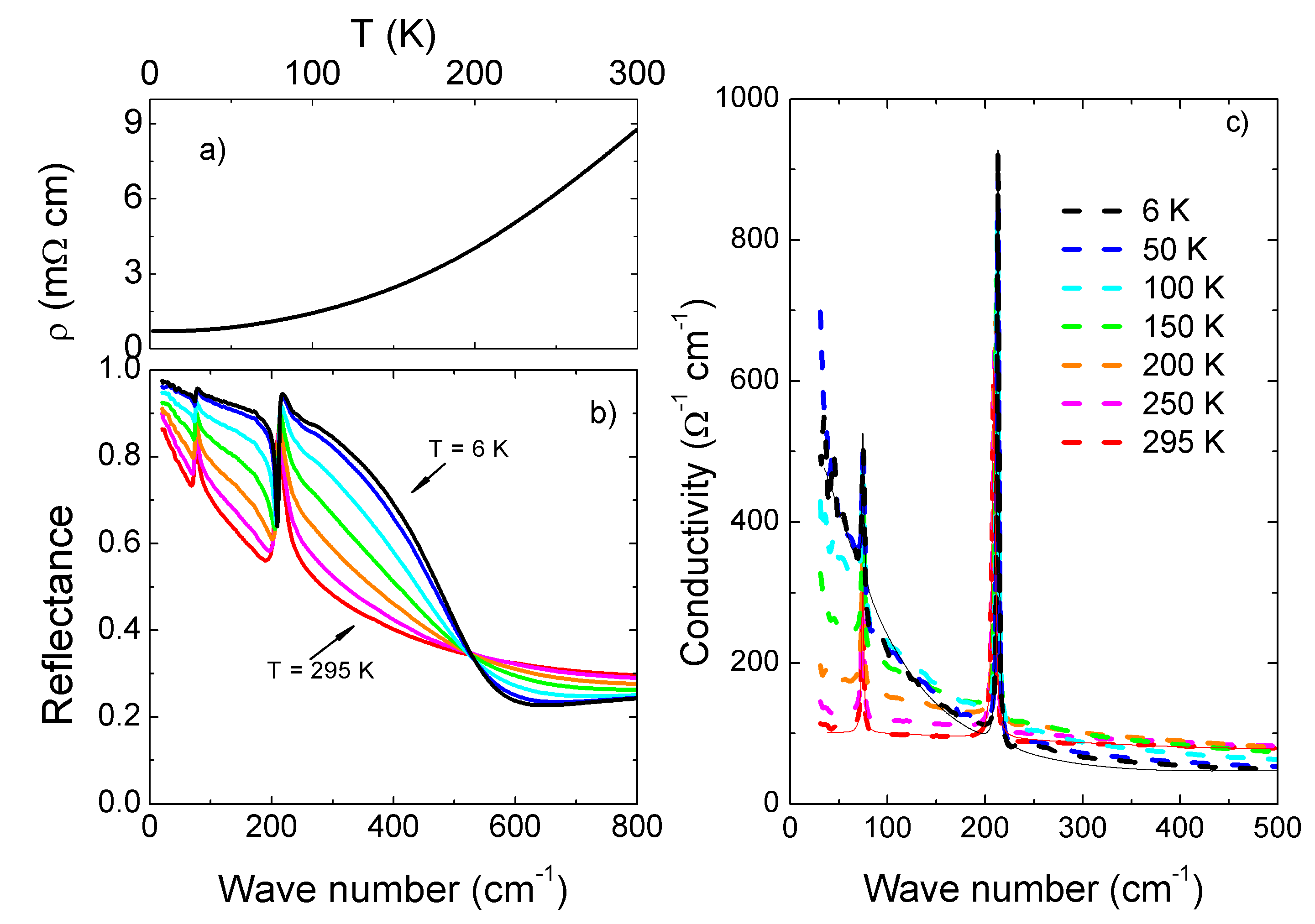}
\caption[]{(Color online) (a) Temperature dependence of the electrical resistivity measured in the ab plane of a single crystal of \KKBMS (b) Optical reflectance measured in the ab plane on a single crystal of \KKBMS for temperatures indicated in (c). (c) Optical conductivity for different temperatures as indicated. Solid lines are fits to the 6 K and 295 K measurements as described in the main text.}
\label{Fig3}
\end{figure}

The temperature dependence of the electrical resisitivty of 2\% K doped \BMS is presented in Figure~\ref{Fig3}a. A metallic $\rho(T)$ = $\rho_0$ + A$\,$T$^2$ temperature dependence typical of a Fermi liquid is observed, with A = 0.09 $\mu\Omega$ cm K$^{-2}$. The measured T-dependence is similar to that previously reported~\cite{pandey2012} but our crystals have slightly lower $\rho_0$ = 0.54 m$\Omega$ cm and a larger residual resistivity ratio $\rho$(300 K)/$\rho$(2 K) $\approx$ 12, indicating good sample quality. The optical reflectance R($\omega$) of \KKBMS at different temperatures is presented in Figure~\ref{Fig3}b. A strong temperature dependence is observed with the reflectance increasing as the temperature decreases, consistent with the formation of a metallic state. The real part of the complex optical conductivity $\sigma_1$ was determined from R($\omega$) via a Kramers-Kronig analysis. Below the lowest measured frequency point, the Hagen-Rubens form has been used for the reflectance, $1-R(\omega) \propto \sqrt{\omega}$, and above the highest-measured frequency point the reflectance was assumed to be constant up to $5 \times 10^4$~cm$^{-1}$, above which a free-electron $1/\omega^4$ response was assumed. Figure~\ref{Fig3}c shows $\sigma_1$ at selected temperatures. All the spectra exhibit clear Drude-like responses, expected for metals.

In order to quantitatively analyze the optical data, we fit $\sigma_1$($\omega$) to the Drude-Lorentz model,

\begin{equation}
\sigma_1(\omega) = \frac{2\pi}{Z_0}\left[ \frac{\omega_p^2}{\tau(\omega^2 + \tau^{-2})} + \sum_j \frac{{\gamma_j \omega^2 \Omega_j^2}}{{(\omega_j^2 - \omega^2)^2 + \gamma_j^2 \omega^2}}\right]
\end{equation}

where Z$_0$ = 377 $\Omega$. The first term describes the free-carrier Drude response, characterized by the plasma frequency $\omega_p$ = 4$\pi$ne$^2$/m*, where n is the carrier concentration and m* is an effective mass, and a scattering rate 1/$\tau$. The second term corresponds to a sum of Lorentz oscillators characterized by a resonance frequency $\omega_j$, a linewidth $\gamma_j$, and an oscillator strength $\Omega_j$.

We find that a single Drude term, which we assign to the doped holes introduced by K, is sufficient to describe $\sigma_1$ (Figure \ref{Fig3}c). This is consistent with the observation that the ordered magnetic moment associated with Mn is not reduced by K doping~\cite{lamsal2013} but remains localized while the observed metallic behavior is due to the doped holes introduced by the K, as previously emphasized~\cite{pandey2013, yeninas2013, bao2012, ueland2015}. These results are in sharp contrast to Ba$_{1-x}$K$_x$Fe$_2$As$_2$ (x=0.4), where two Drude terms corresponding to two different types of charge carriers are necessary to describe the optical conductivity, corresponding to multiple hole and electron pockets~\cite{dai2013}. 

\begin{figure}[[htbp!]
\includegraphics[width=8.6 cm]
{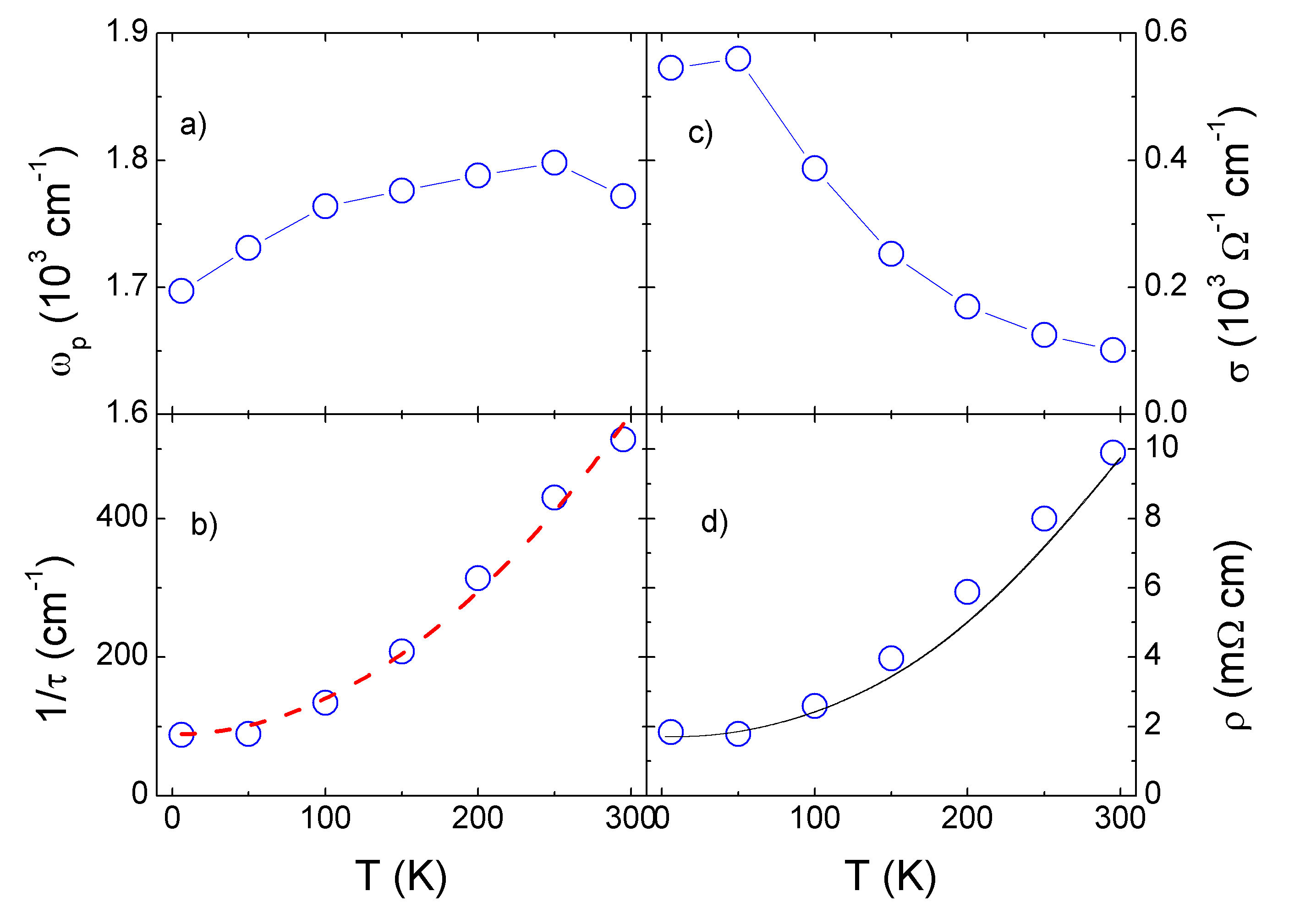}
\caption[]{The temperature dependence of (a) the plasma frequency $\omega_p$, (b) the scattering rate 1/$\tau$, (c) the dc conductivity $\sigma$($\omega$$\rightarrow$0), and (d) the resistivity 1/$\sigma$($\omega$$\rightarrow$0) and the corresponding dc resistivity from transport measurements. For clarity, the dc resistivity is offset by 1 m$\Omega$ cm compared to the same data presented in Figure \ref{Fig3}a}
\label{Fig4}
\end{figure}

Figure \ref{Fig4} shows the temperature dependencies of the Drude parameters taken from our fits. The plasma frequency (Figure \ref{Fig4}a) is almost temperature independent, indicating that the band structure and n/m* do not change appreciably with temperature. The relatively small magnitude of the plasma frequency $\approx$ 1700 cm$^{-1}$ indicates that is a 'bad metal', as expected for only 2\% K doping. This plasma frequency is substantially smaller than in the related compound Ba$_{1-x}$K$_{x}$Fe$_{2}$As$_{2}$ (x = 0.4), where two Drude terms corresponding to two plasma frequencies $\omega_{p1}$ $\approx$ 6000 cm$^{-1}$ and $\omega_{p2}$ $\approx$ 14 000 cm$^{-1}$ are required~\cite{dai2013}. Figure \ref{Fig4}b presents the temperature dependence of the scattering rate of the Drude component, with 1/$\tau$ $\propto$ T$^2$; the dashed line denotes a T$^2$ fit that implies the charge carriers are described by Fermi liquid theory. The temperature dependencies of the dc optical conductivity $\sigma_1$($\omega\rightarrow$0), resisitivity $\rho$ = 1/$\sigma_1$($\omega\rightarrow$0) and dc resisitivity from electronic transport measurements are presented in Figure \ref{Fig4}c,d. Excellent agreement is found between our optical resisitivity and electrical resistivity determined from transport. In both the behavior is well described by Fermi liquid theory, as previously observed and suggested to arise from hole-hole scattering~\cite{pandey2012}. 

The parent state of the iron pnictide superconductors may be regarded as a Hund's metal~\cite{haule2009,georges2013} and it is interesting to ask whether this is also the case for the isostructural Mn pnictide metals. There are two signature features of a Hund's metal: a reduction in the coherence scale for Fermi liquid behavior and the promotion of a metallic state away from half filling. For 2\% K doped \BMS our optical conductivity and transport measurements find the dc resisitivity has a Fermi-liquid like T$^2$ dependence at all measured temperatures. Thus, 2\% K doped \BMS may be viewed as a system of weakly interacting quasiparticles that does not display the unusual non-Fermi liquid properties $\rho$(T)$\propto$T found in the metallic Fe pnictides~\cite{johnston2010}. The optical conductivity of 2\% K doped \BMS is well described by hole conduction corresponding to the introduction of K dopants. This observation, combined with the result that the ordered magnetic moment remains robust in the K-doped metal~\cite{lamsal2013}, suggests the emergence of a metallic state does not promote significant valence fluctuations on the Mn site. Therefore, no signatures characteristic of a possible correlated Hund's metal state have been realized in this system. 

Rather, K doping has transformed \BMS from a local moment antiferromagnetic insulator to a local moment antiferromagnetic metal where the doped holes are responsible for the conduction. The realization of a Hund's metal and superconductivity requires the suppression of magnetic order, as well as metallization. At higher K doping in \BMS itinerant ferromagnetism originating from the As 4p holes has been detected by x-ray magnetic circular dichroism (XMCD)~\cite{pandey2013, ueland2015}. In this report we have shown that BaMn$_2$As$_2$ is very similar to other Mn pnictide systems and we therefore speculate that the ferromagnetism reported in H doped LaMnAsO might not be associated with the Mn site~\cite{hanna2013}. Thus, doping Mn pnictide systems toward metallicity may generically lead to ferromagnetic fluctuations or even order, with limited suppression of the antiferromagnetic order. On the other hand, pressure was found to transform LaMnPO from an antiferromagnetic insulator to an antiferromagnetic metal at 20 GPa and eventually to a paramagnetic metal at 34 GPa~\cite{simonson2012, guo2013} while 20 GPa did not suppress the long range antiferromagnetic order in Ba$_{0.61}$K$_{0.39}$Mn$_2$Bi$_2$ below 300 K~\cite{gu2014}. In both cases, no ferromagnetism was detected and it appears to be the case that only pressure can transform Mn pnictide insulators into potentially correlated Hund's metals, a now familiar breeding ground for superconductivity.


We have reported optical transmission measurements of \BMS that find insulating behavior and a charge gap of 0.86 eV. The measured charge gap is an order of magnitude larger than previously reported and of similar magnitude to that of other Mn pnictide compounds. DFT+DMFT correctly reproduces the charge gap only when Hund's coupling is included. Using these results, we present \BMS as part of a wider class of layered Mn pnictide systems that we classify as Mott-Hund's insulators. We have also presented the optical conductivity of \KKBMS that reveals a Drude peak, characteristic of a metallic state. A Fermi liquid like temperature dependence of $\sigma_1$($\omega$$\rightarrow$0, T) and $\rho(T)$ suggests \KKBMS should not be considered a correlated Hund's metal and thus is an unlikely candidate for high temperature superconductivity. While the outlook for suppressing the local moment antiferromagnetism in these systems by chemical doping remains challenging, the possibility of suppressing the moment by pressure is more positive. We believe a closer examination of the pressure dependence of the magnetic state of layered Mn pnictide compounds, for instance using high pressure neutron scattering techniques, is warranted. 

Research by MCA and CCH at Brookhaven National Laboratory was carried out under the auspices of the US Department of Energy, Office of Basic Energy Sciences, Contract DE-SC0012704. We acknowledge the Office of the Assistant Secretary of Defense for Research and Engineering for providing the National Security Science and Engineering Faculty Fellowship (NSSEFF) funds that supported the research of DEM,SZ,KH,HH, and ZPY. Research was carried out in part at the Center for Functional Nanomaterials at Brookhaven National Laboratory, which is supported by the US Department of Energy , Office of Basic Energy Sciences, under Contract No. DE-SC0012704.


\begin{thebibliography}[
\bibitem{imada1998}M. Imada, A. Fujimori, and Y. Tokura, Rev. Mod. Phys. $\bf{70}$, 1039 (1998).
\bibitem{johnson2015}P.D. Johnson, G. Xu, and W.G. Yin, Iron-Based Superconductivity, (Springer, Switzerland, 2015).
\bibitem{haule2009}K. Haule and G. Kotliar, New J. Phys. $\bf{11}$, 025021 (2009).
\bibitem{georges2013}A. Georges, L. de' Medici, and J. Mravlje, Ann. Rev. Cond. Mat. Phys.  $\bf{4}$, 139 (2013).
\bibitem{mcnally2014}D.E. McNally, J.W. Simonson, K.W. Post, Z.P. Yin, M. Pezzoli, G.J. Smith, V. Leyva, C. Marques, L. DeBeer-Schmitt, A.I. Kolesnikov, Y. Zhao, J.W. Lynn, D.N. Basov, G. Kotliar, and M.C. Aronson, Phys. Rev. B $\bf{90}$, 180403(R) (2014).
\bibitem{mcnally2015}D.E. McNally, J.W. Simonson, J.J. Kistner-Morris, G.J. Smith, J.E. Hassinger, L. DeBeer-Schmitt, A.I. Kolesnikov, I.A. Zaliznyak, and M.C. Aronson, Phys. Rev. B $\bf{91}$, 180407(R) (2015).
\bibitem{singh2009}Y. Singh, A. Ellern, and D.C. Johnston, Phys. Rev. B $\bf{79}$, 094519 (2009).
\bibitem{simonson2012}J. W. Simonson, Z.P. Yin, M. Pezzoli, J. Guo, K. Post, A. Efimenko, N. Hollmann, Z. Hu, H.-J. Lin, C.-T. Chen, C. Marques, V. Leyva, G. Smith, J.W. Lynn, L.L. Sun, G. Kotliar, D.N. Basov, L.H. Tjeng, and M.C. Aronson, Proc. Nat. Acad. Sci. $\bf{109}$, E1815 (2012).
\bibitem{simonson2011} J.W. Simonson, K. Post, C. Marques, G. Smith, O. Khatib, D.N. Basov, and M.C. Aronson, Phys. Rev. B $\bf{84}$, 165129 (2011).
\bibitem{pandey2011}A. Pandey, V.K. Anand, and D.C. Johnston, Phys. Rev. B $\bf{84}$, 014405 (2011).
\bibitem{singh2009_2}Y. Singh, M.A. Green, Q. Huang, A. Kreyssig, R.J. McQueeney, D.C. Johnston, and A.I. Goldman, Phys. Rev. B $\bf{80}$, 100403(R) (2009).
\bibitem{antal2012}A. Antal, T. Knoblauch, Y. Singh, P. Gegenwart, D. Wu, and M. Dressel, Phys. Rev. B $\bf{86}$, 014506 (2012).
\bibitem{kuzmenko2005}A.B. Kuzmenko, Rev. Sci. Instrum. $\bf{76}$, 083108 (2005).
\bibitem{pandey2012}A. Pandey, R.S. Dhaka, J. Lamsal, Y. Lee, V.K. Anand, A. Kreyssig, T.W. Heitmann, R.J. McQueeney, A.I. Goldman, B.N. Harmon, A. Kaminski, and D.C. Johnston, Phys. Rev. Lett. $\bf{108}$, 087005 (2012).
\bibitem{an2009}J. An, A.S. Sefat, D.J. Singh, and Mao-Hua Du, Phys. Rev. B $\bf{79}$, 075120 (2009).
\bibitem{kayanuma2009} Kayanuma,H.Hiramatsu,T. Kamiya, M. Hirano, and H.Hosono, J. Appl. Phys. $\bf{105}$, 073903 (2009).
\bibitem{brechtel1978}E. Brechtel, G. Cordier, and H. Schaefer, Z. Naturforsch. B $\bf{33B}$, 820 (1978).
\bibitem{homes93}C.C. Homes, M. Reedyk, D.A. Crandles, and T. Timusk, Appl. Opt. $\bf{32}$(16), 2976 (1993).
\bibitem{DMFT-RMP2006}G. Kotliar, S.Y. Savrasov, K. Haule, V.S. Oudovenko, O. Parcollet, and C.A. Marianetti, Rev. Mod. Phys. \textbf{78}, 865 (2006).
\bibitem{Haule-DMFT}K. Haule, C.H. Yee and K. Kim, Phys. Rev. B {\bf 81}, 195107 (2010).
\bibitem{wien2k}P. Blaha, K. Schwarz, G. Madsen, D. Kvasnicka, and J. Luitz, (K. Schwarz, Techn. Univ. Wien, 2001).
\bibitem{perdew1996}J. P. Perdew, K. Burke, and M. Ernzerhof, Phys. Rev.Lett. 77, 3865-3868 (1996).
\bibitem{yin2011_2} Z. P. Yin, K. Haule, and G. Kotliar, Nature Materials $\bf{10}$, 932 (2011).
\bibitem{gibson2015}Q.D. Gibson, H. Wu, T. Liang, M.N. Ali, N.P. Ong, Q. Huang, and R.J. Cava, Phys. Rev. B $\bf{91}$, 085128 (2015).
\bibitem{yi2013}M. Yi, D. H. Lu, R. Yu, S. C. Riggs, J.-H. Chu, B. Lv, Z. K. Liu, M. Lu, Y.-T. Cui, M. Hashimoto, S.-K. Mo, Z. Hussain, C. W. Chu, I. R. Fisher, Q. Si, and Z.-X. Shen, Phys. Rev. Lett. $\bf{110}$, 067003 (2013).
\bibitem{brock1994}S.L. Brock, J.E. Greedan, and S.M Kauzlarich, Journal of Solid State Chemistry, $\bf{113}$, 303-311 (1994).
\bibitem{lamsal2013}J. Lamsal, G.S. Tucker, T.W. Heitmann, A. Kreyssig, A. Jesche, Abhishek Pandey, Wei Tian, R.J. McQueeney, D.C. Johnston, and A.I. Goldman, Phys. Rev. B $\bf{87}$, 144418 (2013).
\bibitem{pandey2013}Abhishek Pandey, B.G. Ueland, S. Yeninas, A. Kreyssig, A. Sapkota, Yang Zhao, J.S. Helton, J.W. Lynn, R.J. McQueeney, Y. Furukawa, A.I. Goldman, and D.C. Johnston, Phys. Rev. Lett. $\bf{111}$, 047001, (2013).
\bibitem{yeninas2013}S. Yeninas, Abhishek Pandey, V. Ogloblichev, K. Mikhalev, D.C. Johnston, and Y. Furukawa, Phys. Rev. B $\bf{88}$, 241111(R) (2013).
\bibitem{bao2012}Jin-Ke Bao, Hao Jiang, Yun-Lei Sun, Wen-He Jiao, Chen-Yi Shen, Han-Jie Guo, Ye Chen, Chun-Mu Feng, Hui-Qiu Yuan, Zhu-An Xu, Guang-Han Cao, Ryo Sasaki, Toshiki Tanaka, Kazuyuki Matsubayashi, and Yoshiya Uwatoko, Phys. Rev. B, $\bf{85}$, 144523 (2012).
\bibitem{ueland2015} B.G. Ueland, Abhishek Pandey, Y. Lee, A. Sapkota, Y. Choi, D. Haskel, R.A. Rosenberg, J.C. Lang, B.N. Harmon, D.C. Johnston, A. Kreyssig, and A.I. Goldman, arXiv:1503.07197 (2015).
\bibitem{dai2013}Y.M. Dai, B. Xu, B. Shen, H. Xiao, H.H. Wen, X.G. Qiu, C.C. Homes, and R.P.S.M. Lobo, Phys. Rev. Lett. $\bf{111}$, 117001 (2013).
\bibitem{johnston2010}D.C. Johnston, Advances in Physics 59:6, 803-1061 (2010).
\bibitem{hanna2013}T. Hanna, S. Matsuishi, K. Kodama, T. Otomo, S.I. Shamoto, and H. Hosono, Phys. Rev. B $\bf{87}$, 020401(R) (2013).
\bibitem{guo2013}Jing Guo, J.W. Simonson, Liling Sun, Qi Wu, Peiwen Gao, Chao Zhang, Dachun Gu, Gabriel Kotliar, Meigan Aronson, and Zhongxian Zhao, Nat. Sci. Rep. $\bf{3}$, 2555 (2013).
\bibitem{gu2014}Dachun Gu, Xia Dai, Congcong Le, Liling Sun, Qi Wu, Bayrammurad Saparov, Jing Guo, Peiwen Gao, Shan Zhang, Yazhou Zhou, Chao Zhang, Shifeng Jin, Lun Xiong, Rui Li, Yanchun Li, Xiaodong Li, Jing Liu, Athena S. Sefat, Jiangping Hu, and Zhongxian Zhao, Nat. Sci. Rep. $\bf{4}$, 7342 (2014).

\end{thebibliography}
\end{document}